\documentclass[twocolumn,superscriptaddress,amsmath,amssymb,aps,pra,floatfix]{revtex4-1}

\usepackage[utf8]{inputenc}
\usepackage{wrapfig}
\usepackage{float}
\usepackage{amssymb}
\usepackage{amsfonts}
\usepackage{amsmath}
\usepackage{cases}
\usepackage{stmaryrd}\usepackage{tipa}
\usepackage[dvips]{graphicx}
\usepackage{dcolumn}
\usepackage{bbold}
\usepackage{graphicx}
\usepackage{graphicx}
\usepackage{subfigure}
\usepackage{dcolumn}
\usepackage{bm}
\usepackage{color}
\usepackage{xcolor}
\usepackage{subfigure}
\usepackage{physics}
\usepackage{array}
\usepackage{multirow}
\usepackage{nopageno}
\usepackage{mathtools}
\usepackage{enumerate}
\usepackage[colorlinks=true, linkcolor=red, citecolor=blue]{hyperref}

\begin{document}

\title{Quantum Memristors in Frequency-Entangled Optical Fields}

\author{Tasio Gonzalez-Raya}
\affiliation{Department of Physical Chemistry, University of the Basque Country UPV/EHU, Apartado 644, 48080 Bilbao, Spain}

\author{Joseph M. Lukens}
\affiliation{Quantum Information Science Group, Oak Ridge National Laboratory, Oak Ridge, Tennessee 37831, USA}

\author{Lucas C. C\'{e}leri}
\affiliation{Department of Physical Chemistry, University of the Basque Country UPV/EHU, Apartado 644, 48080 Bilbao, Spain}
\affiliation{Institute of Physics, Federal University of Goi\'{a}s, 74.690-900 Goi\^{a}nia, Goi\'{a}s, Brazil}

\author{Mikel Sanz}
\email{mikel.sanz@ehu.eus}
\affiliation{Department of Physical Chemistry, University of the Basque Country UPV/EHU, Apartado 644, 48080 Bilbao, Spain}

\begin{abstract}
A quantum memristor is a resistive passive circuit element with memory engineered in a given quantum platform. It can be represented by a quantum system coupled to a dissipative environment, in which a system-bath coupling is mediated through a weak measurement scheme and classical feedback on the system. In quantum photonics, such a device can be designed from a beam splitter with tunable reflectivity, which is modified depending on the results of measurements in one of the outgoing beams. Here, we show that a similar implementation can be achieved with frequency-entangled optical fields and a frequency mixer that, working similarly to a beam splitter, produces state superpositions. We show that the characteristic hysteretic behavior of memristors can be reproduced when analyzing the response of the system with respect to the control, for different experimentally-attainable states. Since memory effects in memristors can be exploited for classical and neuromorphic computation, the results presented in this work provides the first steps of a novel route towards constructing quantum neural networks in quantum photonics. 
\end{abstract}

\maketitle
\section{Introduction}
Memory circuit elements are poised to introduce a new paradigm in both classical and quantum computation~\cite{Synap, Neuromorphic, MemLagrangian}. Due to their dependence on previous dynamics, it seems fitting to exploit their passive storage capabilities for enhancement of information processing and for neuromorphic computing tasks. One of these memory circuit elements is called the memristor. It describes a resistive element of an electric circuit that has memory, with a changing resistence whose instantaneous value depends on the history of signals that have crossed the device. This information is codified in the internal variable of the memristor, $\mu$, introducing a state-dependent Ohm's law 
\begin{eqnarray}
\label{eq1}I(t)=G(\mu(t))V(t), \\
\label{eq2}\dot{\mu}(t)=f(\mu(t),V(t)),
\end{eqnarray}
for a voltage-controlled memristor. The dynamic behavior is given by $f(\mu(t),V(t))$, and is manifest in the state-dependent conductance $G(\mu(t))>0$. Attempting to solve Eq.~\ref{eq2} requires time integration over the past of the control signal. This means that the current response given by the voltage-controlled memristor described in Eq.~\ref{eq1} depends, through $G(\mu)$, on previous values of the control voltage, as well as on the present one. Thus, a memristor that undergoes a periodic control signal will display a hysteresis loop when plotting the response versus the control signal (current vs voltage). The slope of this curve is identified with the resistance of the device, and the area enclosed by it is associated with memory effects~\cite{SCQMem}. 

This behavior can de described by Kubo's response theory~\cite{Kubo}, but it was L. O. Chua who, in 1971, coined the term ``memristor'' and described it as an independent element on an electric circuit~\cite{Mem}. It took almost 40 years until such a device was engineered, taking advantage of solid-state electronic and ionic transport properties in nanoscale structures~\cite{HPMem}. Apart from the advantages of using these devices for computation~\cite{AnalogMem}, such as energy efficiency~\cite{Energy}, compared to transistor-based computers, memristors can be also used in machine learning schemes~\cite{PercepMem}. The relevance of the memristor lies in its ubiquitous presence in models which describe natural processes, especially those involving biological systems. For example, memristors inherently describe voltage-dependent ion-channel conductances in the axon membrane in neurons, present in the Hodgkin-Huxley model~\cite{HHM, HHMem}.

The concept of the memristor is a complicated issue to discuss in the quantum realm. The basic proposal consists of a harmonic oscillator coupled to a dissipative environment, where the coupling is changed based on the results of a weak measurement scheme with classical feedback~\cite{QMem}. As a result of the development of quantum platforms in recent years, and their improvement in controllability and scalability, different constructions of a quantum memristor in such platforms have been presented. There is a proposal for implementing it in superconducting circuits~\cite{SCQMem}, exploiting memory effects that naturally arise in Josephson junctions. The second proposal is based on integrated photonics~\cite{OptQMem}: a Mach-Zehnder interferometer can behave as a beam splitter with a tunable reflectivity by introducing a phase in one of the beams, and this is manipulated to study the system as a quantum memristor subject to different quantum state inputs.

In this article, we study a different implementation of a quantum memristor in a quantum photonics setup. Employing beam splitters for frequency-codified quantum states~\cite{Lu2018a}, we explore a new implementation in which the information is codified in frequency-entangled optical fields. We engineer the elements which constitute a quantum memristor, namely, a tunable dissipative element, a weak-measurement scheme, and classical feedback. We find that the characteristic $I$-$V$ curve displays hysteresis loops while subjecting the system to different quantum state inputs. The aim of this work is to establish a building block for memristor-based quantum neural networks in quantum photonics with frequency-codified quantum state inputs, which should have direct applications in quantum machine learning and quantum neural networks~\cite{IntroQML, QML}.

\section{The photonic quantum memristor}
\begin{figure}[h]
{\includegraphics[width=0.35 \textwidth]{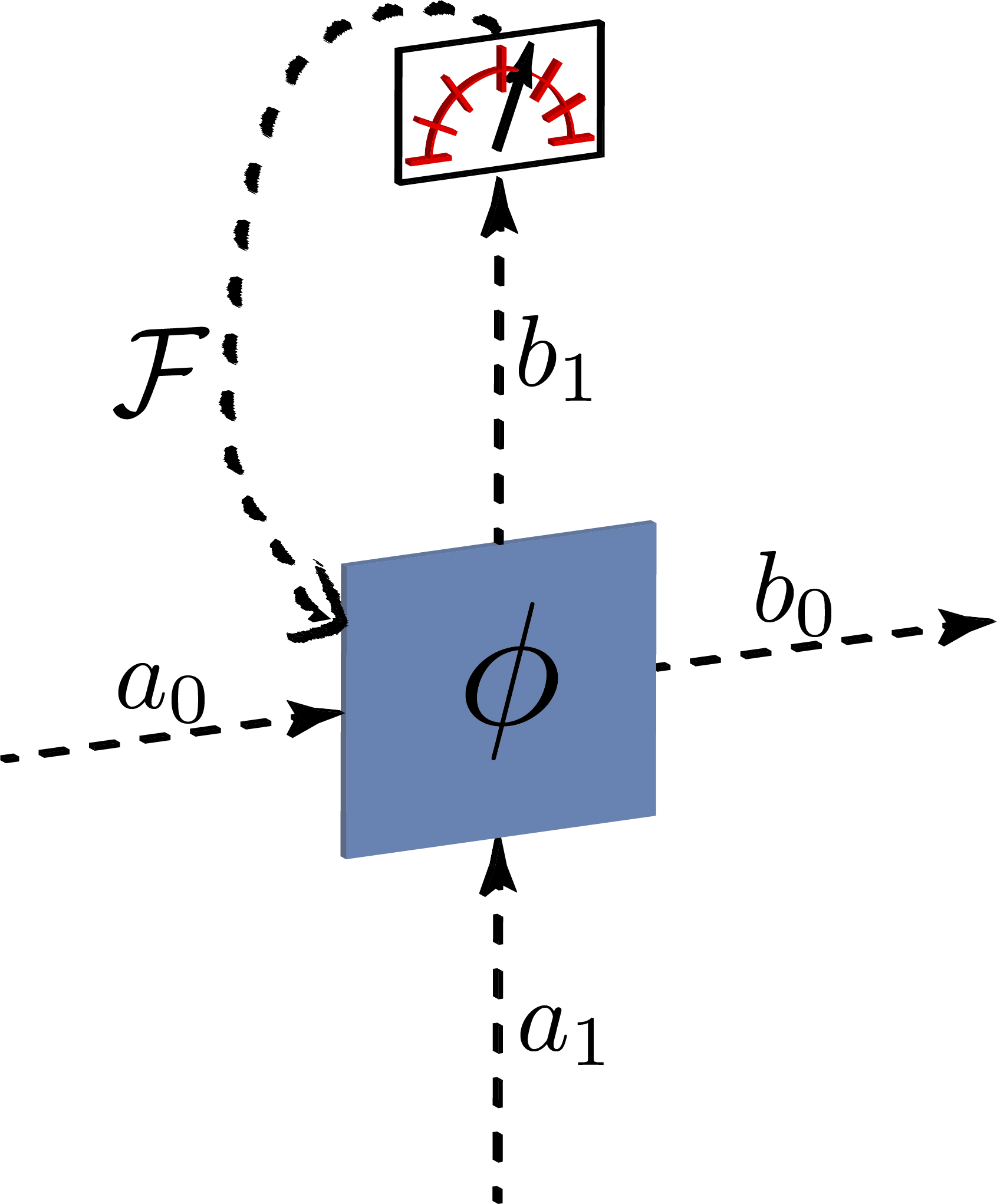}}
\caption{Graphical representation of a photonic quantum memristor. This consists of a beam splitter with tunable reflectivity, a measurement scheme, and a classical feedback system $\mathcal{F}$.}
\label{fig1}
\end{figure}

A memristor can be implemented in quantum optics by means of a beam splitter with a tunable reflectivity. The required non-Markovian dynamics is achieved by inserting a detector in one of the outcomes of the beam splitter (the environment) and, via a feedback mechanism, the reflectivity of the beam splitter is changed. In this way, the coupling between the system (second beam splitter output) and the environment will depend on the previous history of the reflectivity, thus building up memory effects. We present a sketch of this device in Fig. \ref{fig1}.

Formally, a beam splitter with transmitivity $\eta=\cos\phi$ is described by the operator
\begin{equation}\label{eq:beamsplitter}
\hat{B}(\phi,\varphi) = e^{\phi(a^{\dagger}_{0} a_{1}e^{i\varphi}-a_{0} a_{1}^{\dagger}e^{-i\varphi})},
\end{equation}
where $a_0$ and $a_{1}$ are the mode operators for the signal and ancillary signals, respectively, while $\varphi$ stands for an arbitrary fixed phase. The action of the beam splitter on the input modes $\mathbf{a} = (a_0 \,\,\, a_{1})^{T}$ can then be defined as
\begin{equation}
\mathbf{b} = \hat{B}(\phi,\varphi)\mathbf{a},
\end{equation}
with $\mathbf{b} = (b_0 \,\,\, b_{1})^{T}$ being the vector containing the output signal mode $b_0$ and the environmental mode $b_{1}$. The goal of the feedback system $\mathcal{F}$ is to control the value of $\phi$ based on the result of the measurement performed on the environment, thus generating time-correlations, the characteristic feature of non-Markovianity, at the output mode $b_0$. 

\begin{figure}[h]
{\includegraphics[width=0.45 \textwidth]{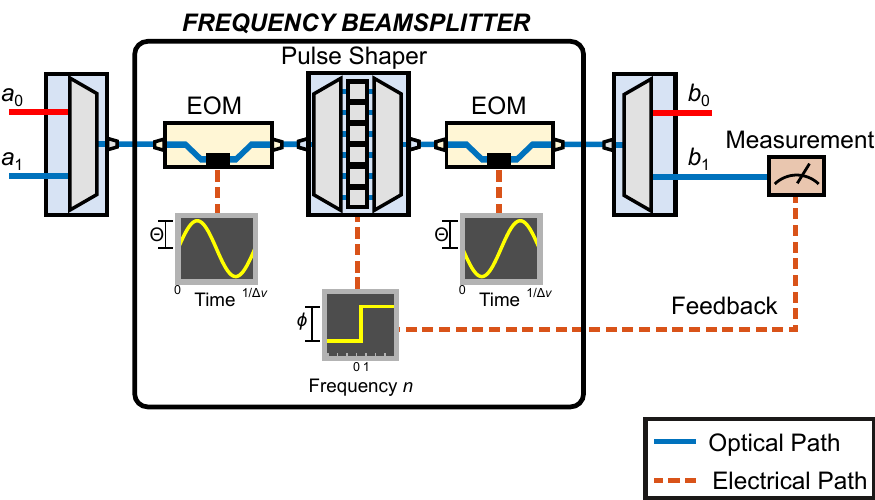}}
\caption{Experimental setup showing a memristor based on the frequency beam splitter.}
\label{fig2}
\end{figure}

We will consider the memristor based on a frequency beam splitter, which is a Hadamard gate acting on the frequency degree of freedom of the input mode, a central element for universal frequency-based quantum computation \cite{Lukens2017}. Recent research has explored implementing unitary operations, such as that in Eq. \ref{eq:beamsplitter}, in time- or frequency-based photonic Hilbert spaces as alternatives to more traditional path encoding approaches. In the frequency-comb-based paradigms, in particular Refs. \cite{Menicucci2008, Lukens2017}, quantum information is carried by photons in discrete modes ($a_n$ or $b_n$), distinguished by index $n\in\mathbb{Z}$ within an equispaced comb defined by frequencies: $\nu_n = \nu_0 + n\Delta\nu$. Such an encoding format proves intriguing given its synergy with fiber-optic networks, applicability to frequency-disparate quantum interconnects, high parallelizability, and compatibility with on-chip photon sources \cite{Kues2019}. However, implementing coherent operations between frequency bins forms a challenging prospect, typically requiring optical nonlinearities mediated by strong pump fields \cite{Kobayashi2016, Clemmen2016}. Yet in 2017, based on electro-optic phase modulators (EOMs) and Fourier-transform pulse shapers, an alternative approach was proposed \cite{Lukens2017}. By cascading EOMs and pulse shapers in an alternating sequence, in principle any frequency-bin unitary can be realized with favorable resource scaling. As these operations are optically linear and precisely controllable, multiple demonstrations have followed this initial proposal, completing the basic pieces of a universal quantum gate set \cite{IEEEptl}.

The 50/50 frequency beam splitter (or Hadamard gate) was the focus of the first experiment in this model \cite{Lu2018a}, where it was found that ---even when restricting to simple, but practically convenient, sinewave-only EOM patterns--- a three-element EOM/pulse shaper/EOM sequence was able to realize a high-fidelity frequency-bin Hadamard with only a slight (2.4\%) reduction in success probability. Further investigation showed that this particular configuration was readily tunable as well; keeping the EOM modulation fixed and modifying only the phase applied by the pulse shaper, the frequency beam splitter reflectivity can be adjusted between 0 and 50\%, a feature exploited for Hong--Ou--Mandel interference in the frequency domain \cite{Lu2018b}. Importantly, this tunability is precisely the prerequisite for a quantum memristor of the form of Fig. \ref{fig1}, thus motivating our detailed exploration of the frequency-bin beam splitter here.

Figure \ref{fig2} furnishes a possible experimental setup for a frequency-bin memristor. Input spectral modes $a$ and $a_a$ are combined into a single fiber where they experience temporal phase modulation at amplitude $\Theta$ and cyclic frequency $\Delta\nu$, followed by a pulse shaper which applies a phase shift $\phi$ to the modes $n \geq 1$ (including those outside of the two-dimensional space of $n\in\{0,1\}$). A second EOM, driven at the same amplitude as the first, but exactly out of phase, concludes the frequency beam splitter. Then the output $b_\mathcal{E}$ is extracted and measured, the results of which are used to update the pulse shaper phase shift $\phi$.

\begin{figure}[t]
{\includegraphics[width=0.5 \textwidth]{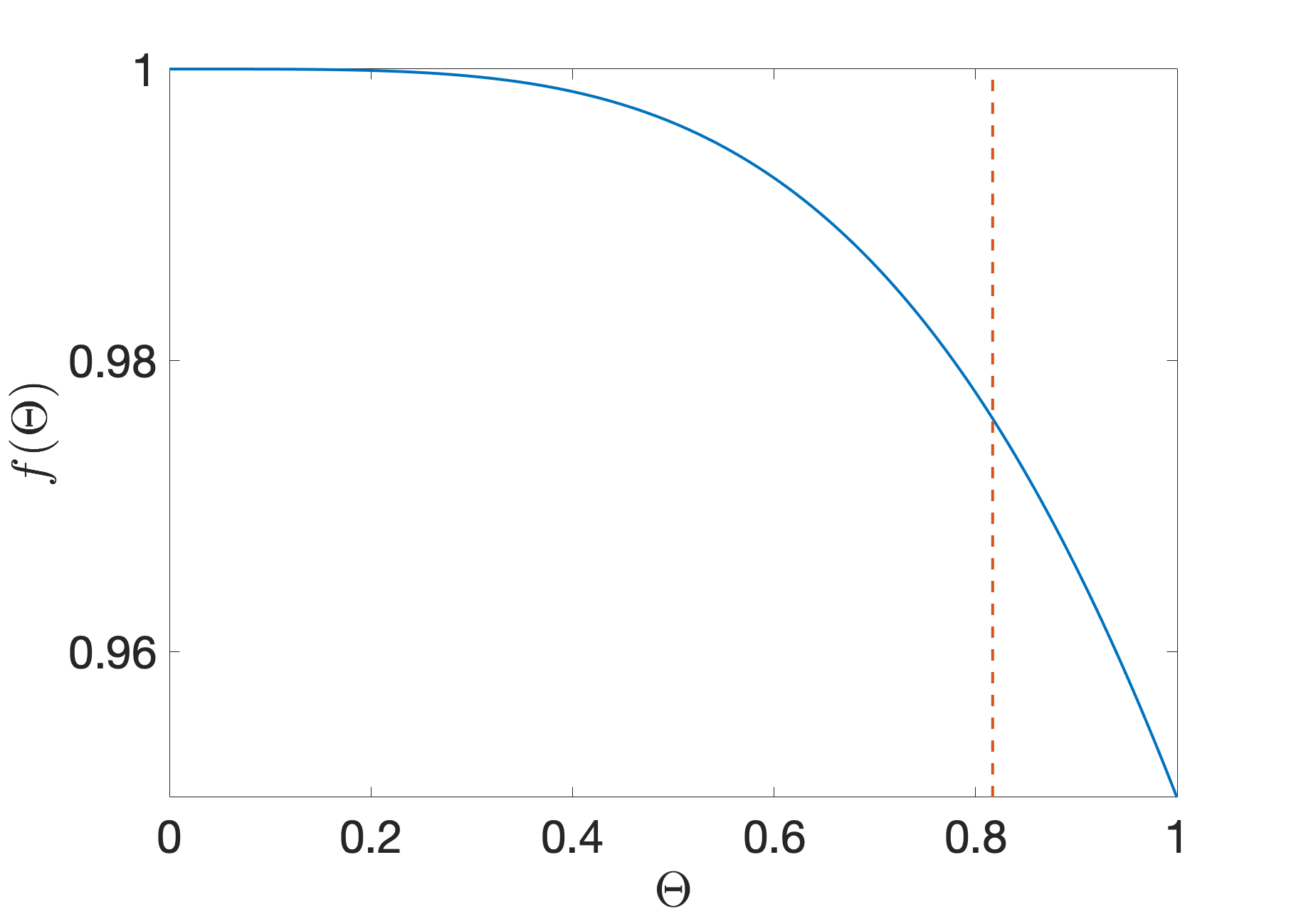}}
\caption{Unitarity of the operator $B$ for any value of the phase $\phi$. The dashed vertical line shows the value $\Theta=0.8169$, which, together with $\phi=\pi$, corresponds to the experimental setup for the realization of the Hadamard gate on such a device.}
\label{fig3}
\end{figure}

Mathematically, each EOM multiplies the input field in the time domain by $\exp\left[\pm i\Theta\sin(2\pi\Delta\nu t)\right]$, for the first and second EOM, respectively. Using a Fourier series expansion, this transformation on frequncy-bin operators can be modeled as $b_m = \sum_n c_{m-n} a_n$ with $c_n = (\mp)^n J_n(\Theta)$, with $J_{n}(\cdot)$ denoting the $n$-th order Bessel function of the first kind. The pulse shaper sandwiched between these two EOMs applies the phase $\phi$ to all bins $n\geq 1$ and zero to all $n\leq 0$. Cascading these three operations, then, we arrive at the following matrix elements connecting the input and output modes of the beam splitter ($n\in\{0,1\})$
\begin{eqnarray}
\nonumber && B_{00} = e^{i\frac{\phi}{2}}\left[\cos\frac{\phi}{2} - iJ_{0}^{2}(\Theta)\sin\frac{\phi}{2}\right], \\
\nonumber && B_{01} = -2i\left[ \sum_{k=1}^{\infty} J_{k}(\Theta)J_{k-1}(\Theta)\right]e^{i\frac{\phi}{2}}\sin\frac{\phi}{2},\\
\nonumber && B_{10} = -2i\left[ \sum_{k=1}^{\infty} J_{k}(\Theta)J_{k-1}(\Theta)\right]e^{i\frac{\phi}{2}}\sin\frac{\phi}{2},\\
&& B_{11} = e^{i\frac{\phi}{2}}\left[\cos\frac{\phi}{2} + iJ_{0}^{2}(\Theta)\sin\frac{\phi}{2}\right].
\label{eq:freqoper}
\end{eqnarray}
The 50/50 beam splitter corresponds to the case $\Theta=0.8169$ and $\phi=\pi$; under these settings, the $2\times 2$ matrix matches the Hadamard operation with fidelity $\mathcal{F}=0.9999$, up to an overall scaling factor that makes the matrix slightly nonunitary due to residual photon scattering into the frequency bins outside of the two-dimensional space~\cite{Lu2018a}. This deviation from unitarity at $\phi=\pi$ can be quantified by $B^\dagger(\Theta,\pi) B(\Theta,\pi) \equiv f(\Theta) \cdot \mathbb{1}$, where
\begin{equation}
f(\Theta) = J_{0}^{4}(\Theta) + 4\left[\sum_{k=1}^{\infty} J_{k}(\Theta)J_{k-1}(\Theta)\right]^{2}.
\end{equation}
Figure \ref{fig3} shows the region where $f(\Theta)$ is close to the value of one required for unitarity; $f(0.8169) = 0.9760$, which corresponds to the success probability $\mathcal{P}$ defined in Ref.~\cite{Lu2018a}. This solution can be viewed as the ``most unitary'' high-fidelity ($\mathcal{F}\geq 0.9999)$ approximation to the Hadamard with two EOMs, one pulse shaper, and single-tone electro-optic modulation. We do note, however, that a fully unitary Hadamard could be realized either by considering arbitrary modulation patterns or adding components; the current settings represent an experimentally valuable compromise between performance and complexity. And so, with $\Theta$ fixed at $0.8169$, adjusting $\phi$ enables tuning of the frequency-bin reflectivity ($|B_{01}|^2=|B_{10}|^2$) and transmissivity ($|B_{00}|^2=|B_{11}|^2$) as needed for the memristor.

\begin{figure}[t]
{\includegraphics[width=0.5 \textwidth]{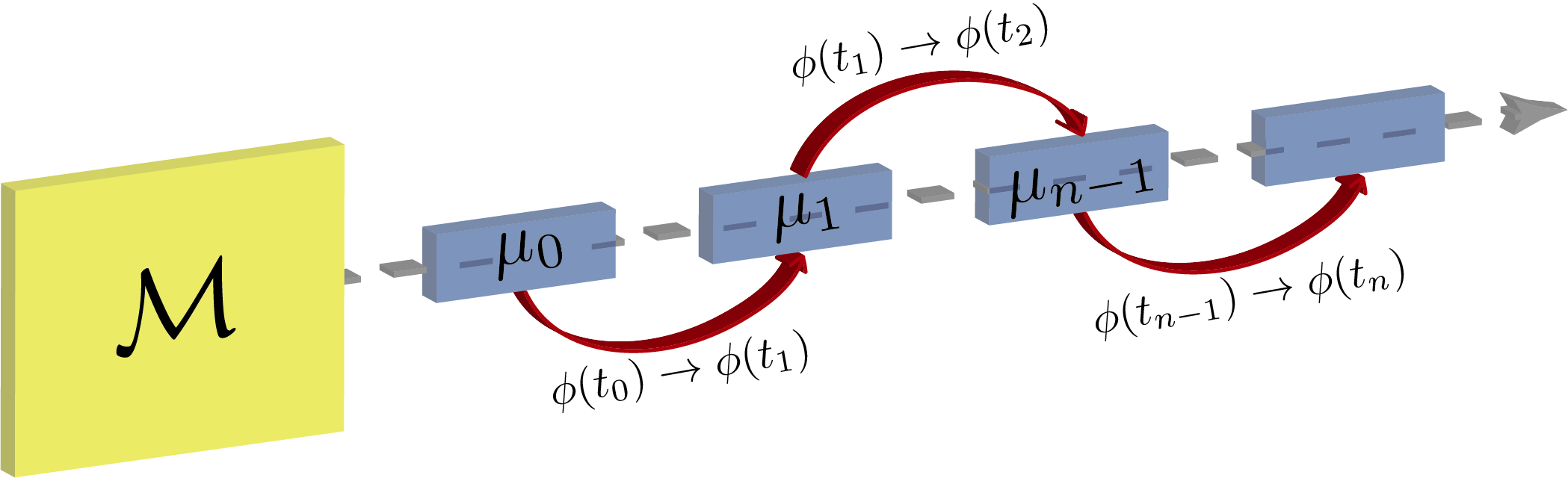}}
\caption{Graphical representation of the measurement schemes that lead to the evolution of the memory variable. Photon number measurements are performed on the environment, followed by a modification of phase $\phi$ depending on the averaged value of these measurements.}
\label{fig4}
\end{figure}

In the measurement and feedback scheme, we aim at modifying the phase $\phi$  appearing in Eq. (\ref{eq:freqoper}) depending on the result of the measurement at the environment output of the beam splitter --- the other output functions as the response signal of the memristor. The measurement scheme is based on photon number measurements, as illustrated in Fig.~\ref{fig4}.

Each $\mu_i$ represents the result of an experiment with a fixed phase, after which we obtain an average of the number of photons in the environment output of the beam splitter, corresponding to the reflected beam. The average number of photons in the outgoing beams, depending on a certain quality of the input state, such as the quadrature $\langle x_{0}^{\text{in}}\rangle$, are 
\begin{eqnarray}
\nonumber && \langle n_{0}^{\text{out}} \rangle = f_{0}(\phi,\langle x_{0}^{\text{in}}\rangle),\\
&& \langle n_{1}^{\text{out}} \rangle = f_{1}(\phi,\langle x_{0}^{\text{in}}\rangle).
\end{eqnarray}
$\phi$ is modified according to the latter, following a dynamic equation
\begin{equation}
\dot{\phi} = g(\phi,\langle n_{1}^{\text{out}}\rangle),
\end{equation}  
which we are free to choose. For illustrative purposes, we propose oscillating input states, for example $\langle x_{0}^{\text{in}}\rangle = \langle x_{0}^{\text{in}}\rangle_{\text{max}}\cos\omega t$, where $\omega$ is a free parameter that we can choose to optimize the correlations and memory persistence in the system. 

The result of the measurement process corresponds to a time average of the number of photons, reflected in the beam splitter, that are detected, with fixed $\phi$, where $\tau_{k}$ defines the duration of one complete experiment $k$. The result of each experiment is then used to update $\phi$, which is changed for time slices between experiments. However, the global change of $\phi$ occurs in a timescale given by $\omega^{-1}$, whereas the duration of each experiment is $\tau_{k}$, so to have a continuous dynamic equation for the update of $\phi$ we are assuming $\tau_{k}\omega \ll 1$. 

The hysteresis loop area, characteristic of the memristor's non-Markovian behavior, can be understood as a memory quantifier. Our goal here is to optimize the feedback process to obtain maximum time-correlation at the output signal, since this quantity will be related to the memory persistence in the system. This is crucial, especially when designing a neural network based on quantum memristors, where we would look to minimize decoherence in order to take advantage of quantum effects. This optimization will also allow us to study the persistence of the memory effects in the system, affected too by the feedback mechanism. 

Now, let us look into the effect of this device on different initial states, analyzing their hysteretic response.

\subsection{Coherent states}
The two-mode coherent states are given by
\begin{equation}
|\alpha_{\omega_{0}},\beta_{\omega_{1}}\rangle = e^{-(|\alpha|^{2}+|\beta|^{2})/2}\sum_{n,m=0}^{\infty}\frac{\alpha^{n}\beta^{m}}{\sqrt{n! m!}} (a^{\dagger}_{\omega_{0}})^{n}(a^{\dagger}_{\omega_{1}})^{m}|0,0\rangle,
\end{equation}
where $\alpha$ and $\beta$ are complex numbers that can be experimentally tuned. 

Remember that a coherent state $|\alpha\rangle$ can be defined by the displacement operator $D(\alpha)=e^{\alpha a^{\dagger}-\bar{\alpha}a}$ acting on the vaccum, $|\alpha\rangle = D(\alpha)|0\rangle$. Using the equality
\begin{equation}\label{trick}
e^{A}B e^{-A} = \sum_{k=0}^{\infty} \frac{1}{k!}[A,[A, ...,[A,B]...]]
\end{equation}
derived from the Baker-Campbell-Hausdorff transformation for any two operators $A,B$, we arrive at
\begin{equation}
D(\alpha)^{\dagger}a D(\alpha) = a + \alpha. 
\end{equation}
By applying displacement operators $D(\alpha)$, $D(\beta)$ on the first and second beams, respectively, we achieve the following transformation
\begin{equation}
\begin{pmatrix} a_{0} \\  a_{1} \end{pmatrix} \longrightarrow  
\begin{pmatrix} D^{\dagger}(\alpha)a_{0}D(\alpha) \\ D^{\dagger}(\beta)a_{1}D(\beta) \end{pmatrix} = \begin{pmatrix} a_{0} + \alpha \\ a_{1} + \beta \end{pmatrix}
\end{equation}
and these modes are the input to the beam splitter. Then, we compute
\begin{eqnarray}
\nonumber && \begin{pmatrix} b_{0} \\  b_{1} \end{pmatrix}_{\text{coh}} = B(\phi)  \begin{pmatrix} a_{0} + \alpha \\ a_{1} +\beta \end{pmatrix}  =\\
&& \begin{pmatrix} B_{00}(a_{0}+\alpha) + B_{01}(a_{1}+\beta) \\ B_{01}(a_{0}+\alpha) + e^{-i\phi}\bar{B}_{00}(a_{1}+\beta) \end{pmatrix}, 
\end{eqnarray}
where we have identified $B_{11}=e^{-i\phi}\bar{B}_{00}$ and $B_{10}=B_{01}$. Considering a vacuum state in the second ingoing beam ($\beta=0$), we compute the number of photons in the first outgoing beam,
\begin{equation}
\langle n_{0}^{\text{out}}\rangle = \langle 0,0| b_{0}^{\dagger}b_{0} |0,0\rangle = |\alpha|^{2} |B_{00}|^{2} = \langle n_{0}^{\text{in}}\rangle |B_{00}|^{2}.
\end{equation}
See that $\langle n_{0}^{\text{in}}\rangle=\langle x_{0}^{\text{in}}\rangle^{2}$ for $\alpha\in\mathbb{R}$, assuming a displacement in the $x$-direction, where $x = \frac{a+a^{\dagger}}{2}$ is the quadrature operator. Consider that the response of the system is codified in $\langle n_{0}^{\text{out}}\rangle$; this implies that the measured quantity will be $\langle n_{1}^{\text{out}}\rangle = |\alpha|^{2} |B_{01}|^{2}$. 

\begin{figure}[t]
{\includegraphics[width=0.5 \textwidth]{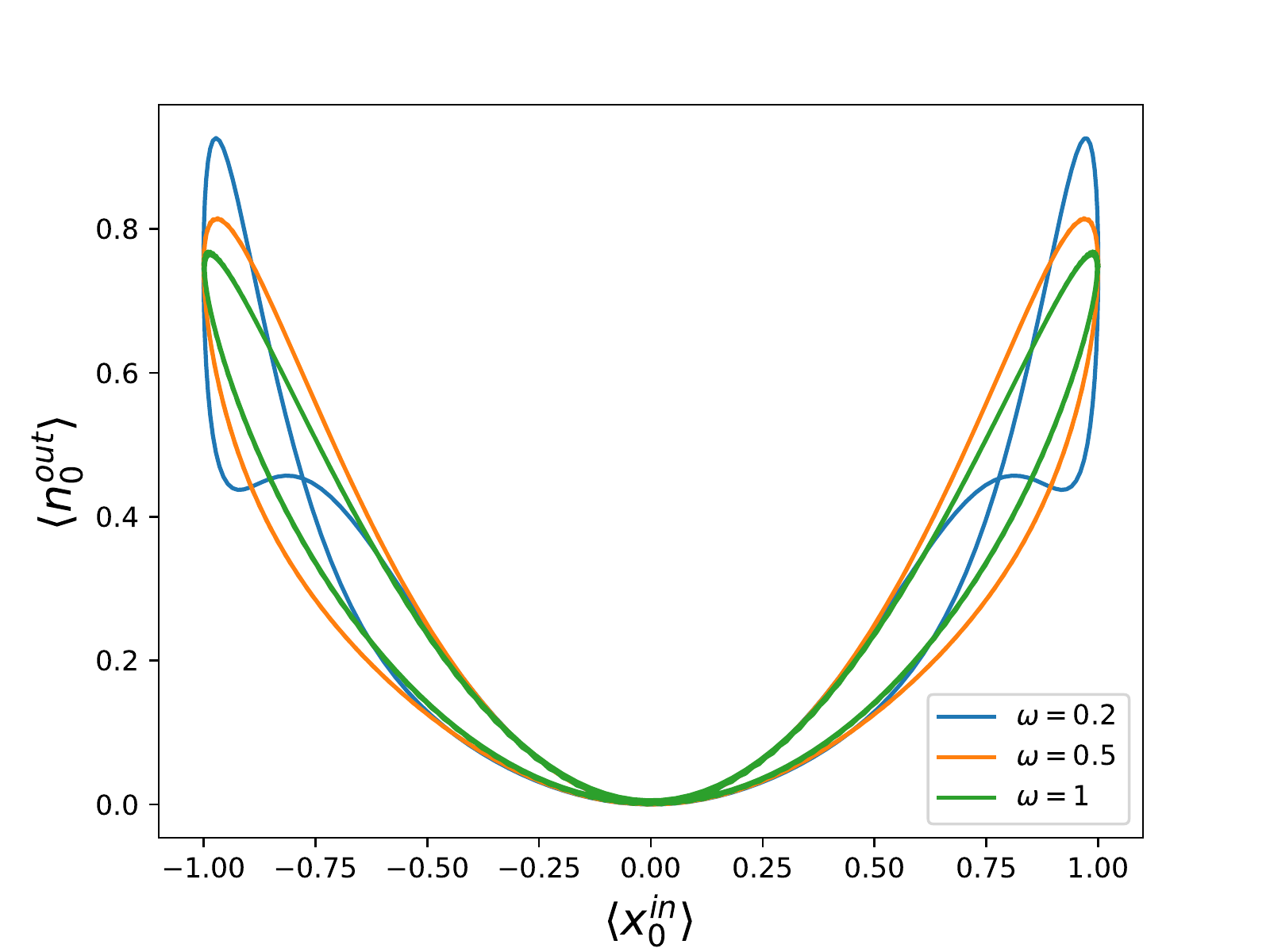}}
\caption{Number of photons $\langle n_{0}^{\text{out}}\rangle$ of the outgoing beam versus the quadrature of the input state $\langle x_{0}^{\text{in}}\rangle$ for coherent states, displaying a pinched hysteresis loop, proving that this system behaves as a memristor. We have plotted this for three different frequencies: $\omega=0.2$ (blue), $\omega=0.5$ (orange), and $\omega=1$ (green), exemplifying  that the area of the loop decreases for higher frequencies. We have used $\phi(0)=\frac{\pi}{2}$, $\omega_{0}=1$, and $\langle x_{0}^{\text{in}}\rangle_{\text{max}}=1$.}
\label{fig5}
\end{figure}

Having identified the response and the internal variable, we can write the equations of the memristor,
\begin{eqnarray}
\nonumber \langle n_{0}^{\text{out}}\rangle &=& f(\phi,\langle x_{0}^{\text{in}}\rangle) \langle x_{0}^{\text{in}}\rangle, \\
\dot{\phi} &=& g(\phi,\langle x_{0}^{\text{in}}\rangle)
\end{eqnarray}
Since we have freedom to choose the update of the phase $\phi$, we propose a simple function for illustrative purposes,
\begin{equation}
\dot{\phi} = \frac{\omega_{0}}{\langle x_{0}^{\text{in}}\rangle_{\text{max}}}\langle x_{0}^{\text{in}}\rangle.
\end{equation}
Assuming that we are able to pump the system to induce periodicity in the quadrature of the input state, such that $\langle x_{0}^{\text{in}}\rangle = \langle x_{0}^{\text{in}}\rangle_{\text{max}} \cos\omega t$, the evolution of $\phi$ is described by 
\begin{equation}
\phi(t) = \phi(0) + \frac{\omega_{0}}{\omega} \sin\omega t.
\end{equation}
This implies that
\begin{eqnarray}
\nonumber f(\phi,\langle x_{0}^{\text{in}}\rangle) &=& \langle x_{0}^{\text{in}}\rangle_{\text{max}} \left[ \cos^{2}\frac{\phi(t)}{2} + J_0(\Theta)^{4}\sin^{2}\frac{\phi(t)}{2} \right]\cos\omega t ,\\
g(\phi,\langle x_{0}^{\text{in}}\rangle) &=& \omega_{0} \cos\omega t.
\end{eqnarray} 
In Fig.~\ref{fig5}, we represent $\langle n_{0}^{\text{out}}\rangle$ versus $\langle x_{0}^{\text{in}}\rangle$ to observe hysteretic behavior, related to that appearing in the I-V characteristic curve of memristors. The hysteresis loop in this case is pinched, and its area decreases with an increasing frequency of the driving, which means that this system behaves as a memristor in these variables.

\subsection{Squeezed states}
\begin{figure}[t]
{\includegraphics[width=0.5 \textwidth]{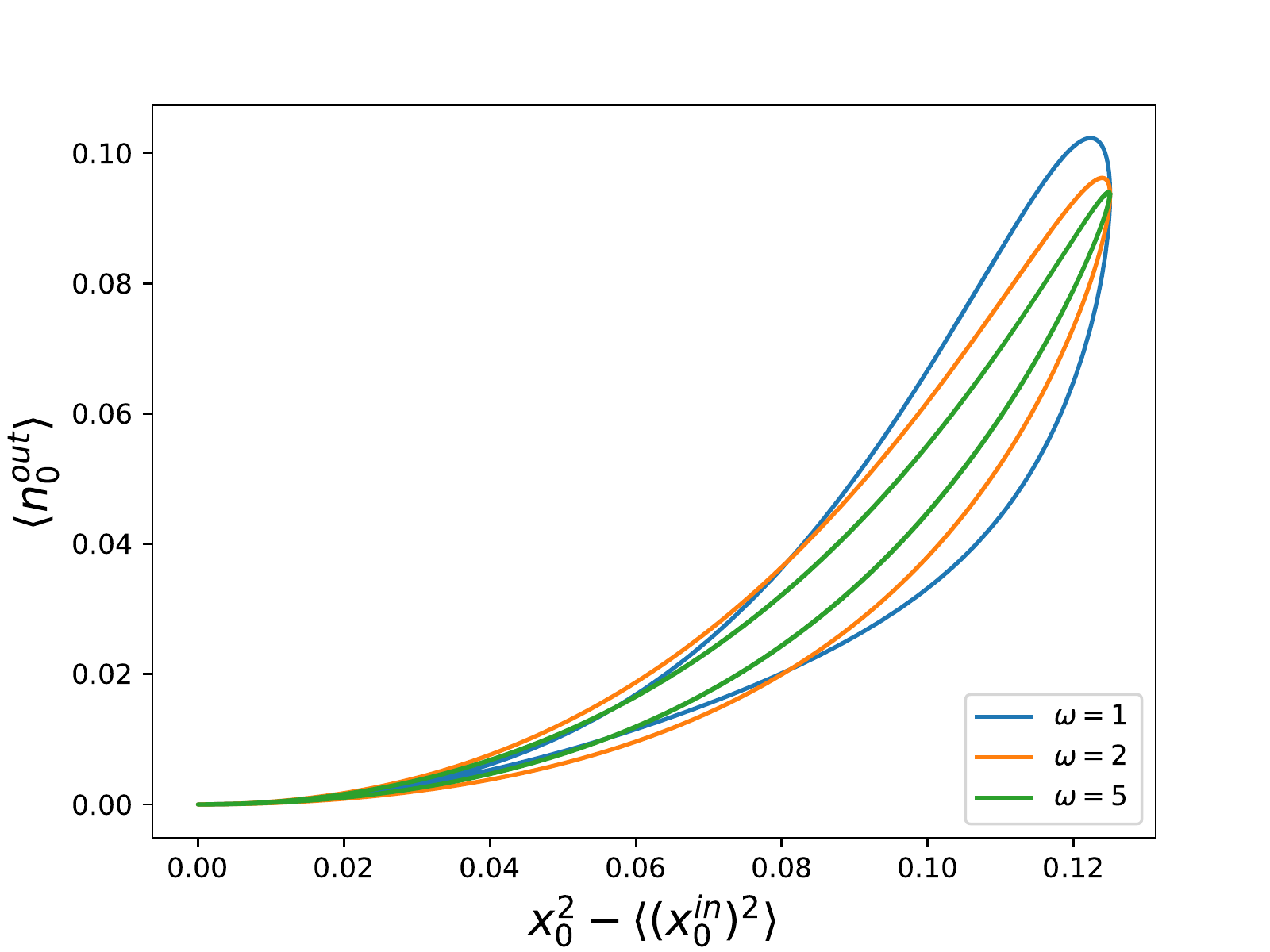}}
\caption{Number of photons $\langle n_{0}^{\text{out}}\rangle$ of the outgoing beam versus the second-moment of the quadrature of the input state $\langle (x_{0}^{\text{in}})^{2} \rangle$ for squeezed states, displaying a pinched hysteresis loop, proving that this system behaves as a memristor. We have plotted this for three different frequencies: $\omega=1$ (blue), $\omega=2$ (orange), and $\omega=5$ (green), exemplifying  that the area of the loop decreases for higher frequencies. We have used $\phi(0)=\frac{\pi}{2}$, $\omega_{0}=5$, $\alpha = 0.5$, and $x_0 = 1/4$.}
\label{fig6}
\end{figure}

It is interesting to study the response of the system when considering squeezed state inputs. Analogous to the displacement operator for coherent states, we can define the squeezing operator $S(z) = e^{\frac{1}{2}(\bar{z}a^{2}-z a^{\dagger 2})}$, with $z=re^{i\varphi}$, such that a squeezed state is defined as $|z\rangle = S(z)|0\rangle$. Using the relation in Eq.~\ref{trick}, we can define the transformation
\begin{equation}
S^{\dagger}(z)aS(z) = a \cosh r - e^{i\varphi} a^{\dagger} \sinh r.
\end{equation}
By applying squeezing operators $S(z_0), S(z_1)$ on the first and second beam, respectively, we obtain
\begin{equation}
\begin{pmatrix} a_{0} \\  a_{1} \end{pmatrix} \rightarrow  
\begin{pmatrix} S^{\dagger}(z_0)a_{0}S(z_0) \\ S^{\dagger}(z_1)a_{1}S(z_1) \end{pmatrix} = \begin{pmatrix} a_0 \cosh r_0 - e^{i\varphi_0} a_0^{\dagger} \sinh r_0 \\ a_! \cosh r_1 - e^{i\varphi_1} a_1^{\dagger} \sinh r_1 \end{pmatrix}
\end{equation}
which represent the inputs to the beam splitter. These modes are modified by the beam splitter as follows,
\begin{eqnarray}
\nonumber && \begin{pmatrix} b_{0} \\  b_{1} \end{pmatrix}_{\text{squ}} = B(\phi)  \begin{pmatrix} a_0 \cosh r_0 - e^{i\varphi_0} a_0^{\dagger} \sinh r_0 \\ a_1 \cosh r_1 - e^{i\varphi_1} a_1^{\dagger} \sinh r_1 \end{pmatrix}. 
\end{eqnarray}
We consider a vacuum state in the second ingoing beam ($r_1 = 0$), and compute the number of photons in the first outgoing beam,
\begin{equation}
\langle n_{0}^{\text{out}}\rangle = \langle 0,0| b_{0}^{\dagger}b_{0} |0,0\rangle = \sinh^{2} r_0 |B_{00}|^{2}
\end{equation}
as the response of the system. For the control variable, we choose $\langle x^{2} \rangle = \frac{1}{4} \langle (a + a^{\dagger})^{2}\rangle$ for the first ingoing beam
\begin{equation}
\langle (x_0^{\text{in}})^{2}\rangle = \frac{1}{4}(\cosh^{2}r + \sinh^{2}r - \sinh 2r \cos\varphi)
\end{equation}
where we have set $r_0 = r$ and $\varphi_0 = \varphi$. In this setup, our goal is to identify a memristive system satisfying the following equations,
\begin{eqnarray}
\nonumber \langle n_{0}^{\text{out}}\rangle &=& f(\phi,\langle (x_0^{\text{in}})^{2}\rangle) \langle (x_0^{\text{in}})^{2}\rangle, \\
\dot{\phi} &=& g(\phi, \langle (x_0^{\text{in}})^{2}\rangle),
\end{eqnarray}
with the phase $\phi$ set as the memory variable, as in the previous section. Fixing the squeezing in the x axis ($\varphi=0$), we can write 
\begin{equation}
\langle (x_0^{\text{in}})^{2}\rangle = \frac{1}{4}(\cosh r - \sinh r)^{2},
\end{equation}
and so
\begin{equation}
1-4\langle (x_0^{\text{in}})^{2}\rangle =2\sinh r (\cosh r - \sinh r).
\end{equation}
Then, we can write
\begin{equation}
f(\phi,\langle (x_0^{\text{in}})^{2}\rangle) = \left[ \frac{1-4\langle (x_0^{\text{in}})^{2}\rangle}{4\langle (x_0^{\text{in}})^{2}\rangle}\right]^{2} |B_{00}|^{2}.
\end{equation}
The number of photons measured in the outgoing beam corresponding to the environment is given by
\begin{equation}
\langle n_{1}^{\text{out}}\rangle = 16 \left[ \frac{x_{0}^{2} -  \langle (x_0^{\text{in}})^{2}\rangle}{ \langle (x_0^{\text{in}})^{2}\rangle}\right]^{2} |B_{01}|^{2}  \langle (x_0^{\text{in}})^{2}\rangle,
\end{equation}
from which $\langle (x_0^{\text{in}})^{2}\rangle$ can be obtained. As the update of the memory variable, we propose the function
\begin{equation}
\dot{\phi} = \pm \frac{\omega_0}{x_0} \sqrt{x_0^{2} - \langle(x_0^{\text{in}})^{2}\rangle},
\end{equation}
where $x_0 ^{2}= \langle (x_0^{\text{in}})^{2}\rangle_{\text{vac}} = 1/4$. Assuming we are able to engineer a periodic pumping $\langle (x_0^{\text{in}})^{2}\rangle = (1-\alpha\cos^{2}\omega t)/4$ for the input states, we have that
\begin{equation}
g(\phi , \langle (x_0^{\text{in}})^{2}\rangle) = \pm \omega_{0} \sqrt{\alpha}\cos\omega t,
\end{equation}   
leading to the evolution of $\phi$ to be given by
\begin{equation}
\phi(t) = \phi(0) + \frac{\omega_0}{\omega}\sqrt{\alpha}\sin\omega t.
\end{equation}
We can observe hysteresis loops when representing the number of photons in the first outgoing beam versus $x^{2}$ in the input beam, as can be seen in Fig.~\ref{fig6} 

\begin{figure}[t]
{\includegraphics[width=0.5 \textwidth]{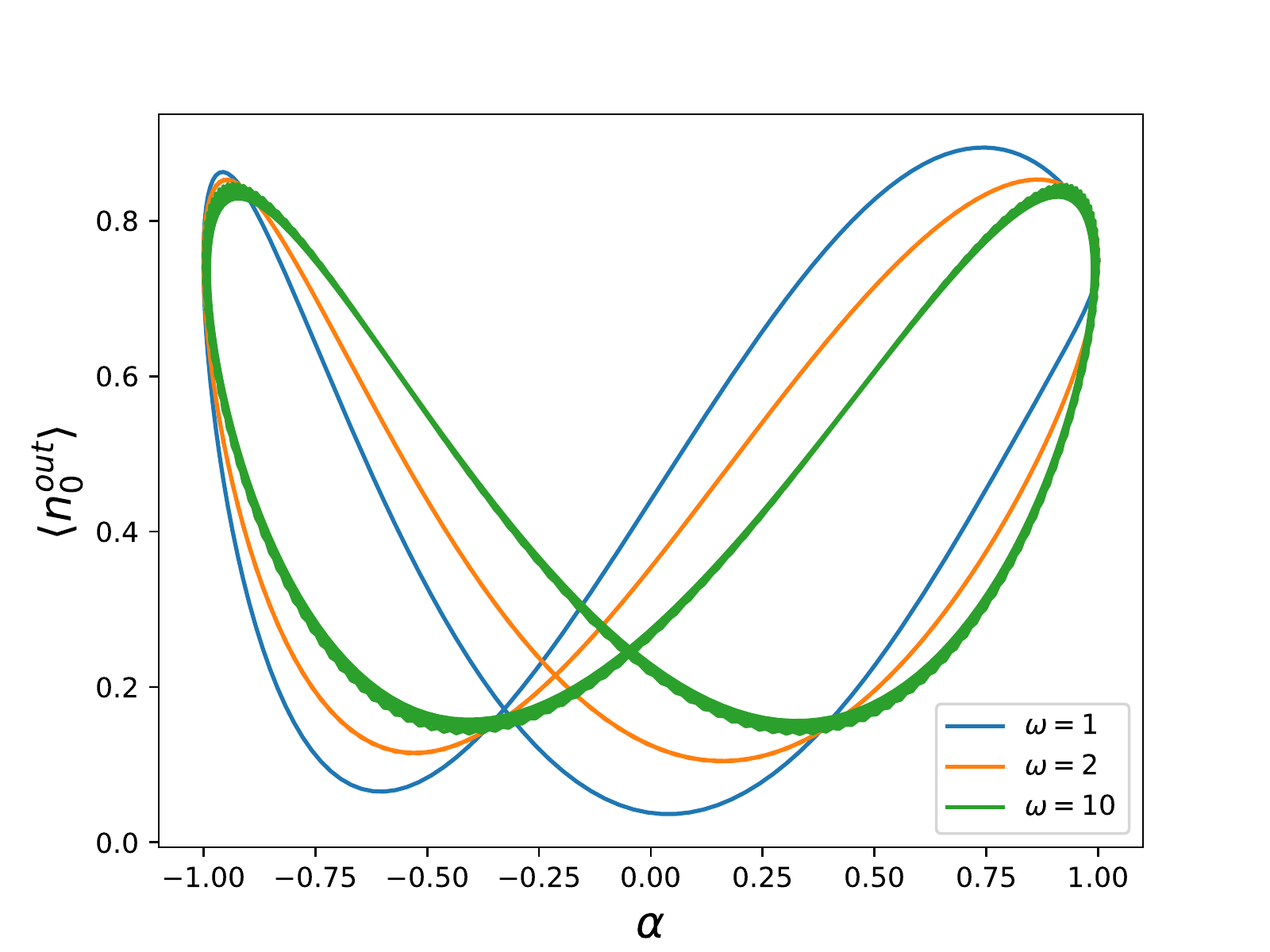}}
\caption{Number of photons $\langle n_{0}^{\text{out}}\rangle$ of the outgoing beam versus $\alpha=\cos\omega t$ for entangled fock states, displaying a pinched hysteresis loop,where the point of crossing approaches zero with higher $\omega$. This system does not behave as a memristor. We have plotted this for three different frequencies: $\omega=1$ (blue), $\omega=2$ (orange), and $\omega=10$ (green), exemplifying that the area of the loop does not seem to decrease for higher frequencies. We have used $\phi(0)=\frac{\pi}{2}$, and $\omega_{0}=1$.}
\label{fig7}
\end{figure}

\subsection{Fock states}
In this category, we can use a variety of two-photon Fock states, among the ones below,
\begin{eqnarray}
\nonumber |\psi_{1}\rangle = \alpha|1_{\omega_{0}}, 0_{\omega_{1}}\rangle + \beta|0_{\omega_{0}}, 1_{\omega_{1}}\rangle &=& (\alpha a^{\dagger}_{\omega_{0}}+\beta a^{\dagger}_{\omega_{1}})|0,0\rangle,\\
\nonumber |\psi_{2}\rangle = |1_{\omega_{0}}, 1_{\omega_{1}}\rangle &=& a^{\dagger}_{\omega_{0}}a^{\dagger}_{\omega_{1}}|0,0\rangle,\\
\nonumber |\psi_{3}\rangle = |2_{\omega_{0}}, 0_{\omega_{1}}\rangle &=& \frac{1}{\sqrt{2}}(a^{\dagger}_{\omega_{0}})^{2}|0,0\rangle,\\
|\psi_{4}\rangle = |0_{\omega_{0}}, 2_{\omega_{1}}\rangle &=& \frac{1}{\sqrt{2}}(a^{\dagger}_{\omega_{1}})^{2}|0,0\rangle.
\end{eqnarray}
The only Fock input state, among the ones given above, that allows for a change in the control over the timescale of the feedback mechanism is the first one, $|\psi_{1}\rangle$. We begin from this state and compute the number of photons in the outgoing beam,
\begin{eqnarray}
\nonumber &&\langle n_{0}^{\text{out}}\rangle_{\psi_{1}} = \langle\psi_{1}|b_{0}^{\dagger}b_{0}|\psi_{1}\rangle = \\
&& \langle\psi_{1}| (\bar{B}_{00}a^{\dagger}_{0} + \bar{B}_{01}a^{\dagger}_{1})(B_{00}a_{0} + B_{01}a_{1}) |\psi_{1}\rangle,
\end{eqnarray}
and choose $\alpha=\cos\omega t$ and $\beta=\sin\omega t$ to obtain
\begin{equation}
\langle n_{0}^{\text{out}}\rangle_{\psi_{1}} = \cos^{2}\omega t \cos^{2}\frac{\phi}{2} + (c_{1}\cos\omega t + 2c_{2}\sin\omega t)^{2}\sin^{2}\frac{\phi}{2},
\end{equation}
where $c_{1} = J_{0}^{2}(\Theta)$ and $c_{2}=\sum_{k=1}^{\infty}J_{k}(\Theta)J_{k-1}(\Theta)$. The number of photons dissipated to the environment is then 
\begin{equation}
\langle n_{1}^{\text{out}}\rangle_{\psi_{1}} = \sin^{2}\omega t \cos^{2}\frac{\phi}{2} + (c_{1}\sin\omega t + 2c_{2}\cos\omega t)^{2}\sin^{2}\frac{\phi}{2}.
\end{equation}
From these measurements, $\alpha$ can be inferred to design the following update of the memory variable,
\begin{equation}
\dot{\phi} = \omega_{0}\alpha = \omega_{0}\cos\omega t,
\end{equation}
such that its time evolution is described by
\begin{equation}
\phi(t) = \phi(0) + \frac{\omega_{0}}{\omega} \sin\omega t.
\end{equation}
In Fig.~\ref{fig7} we represent this photon number against $\alpha$, which represents the square root of the photon number in the first ingoing beam,
\begin{equation}
\langle n_{0}^{\text{in}}\rangle_{\psi_{1}} = \langle\psi_{1}| a_{0}^{\dagger}a_{0}|\psi_{1}\rangle = |\alpha|^{2}.
\end{equation}
We obtain hysteretic behavior in the system, but the loops cross at a point that is moving with the frequency $\omega$ of the driving, approaching $0$ for higher frequencies, and the area of the loops does not seem to decrease with increasing frequency. Since the system and the environment are entangled, this system does not represent a memristor.

\section{Discussions}
We have followed a scheme to build a resistive memory element in a quantum platform. We have reproduced this behavior in quantum photonics with frequency-codified quantum states, by engineering a frequency mixer as a tunable beam splitter with a measurement scheme that modifies its reflectivity through classical feedback. Hysteretic behavior was found when representing the response of the system versus the control, a sign of memristive systems. The scalability of such devices can be tested to construct quantum neural networks in a quantum photonics platform, which can represent a direct hardware-based implementation of quantum machine learning algorithms.

In this study, we have focused on a frequency-bin memristor design that is feasible. The tunable beam splitter outlined in Fig.~\ref{fig2} and expressed by Eq.~(\ref{eq:freqoper}) has been experimentally demonstrated with behavior matching theory extremely well~\cite{Lu2018a, Lu2018b}. Nevertheless, practical limitations present challenges toward realizing this memristor's full potential in the laboratory. For example, in the tabletop demonstrations of the frequency beam splitter thus far, component insertion losses have led to overall throughputs $\sim$5\%---a significant limitation, particularly for continuous-variable encoding, and much lower than the the $\mathcal{P}=97.6\%$ indicated in the lossless theory. Nonetheless, unlike the success probability $\mathcal{P}$, insertion losses stem from nonidealities (e.g., mode mismatch, waveguide loss) that can in principle be eliminated through device engineering. In fact, integrated microring-based pulse shapers~\cite{Wang2015} from existing foundries, coupled with ultralow-loss EOMs~\cite{Wang2018}, provide a promising outlook for chip-scale frequency memristors with markedly lower loss.

Additionally, our memristor design relies heavily on real-time feedback of the phase shift $\phi$. Ideally, given mode separation $\Delta\nu\approx 25$~GHz, one would like update speeds in the $\sim$ns regime---fast, but sufficiently slower than the RF period, in order to retain the validity of the frequency-bin model. Such refresh rates are beyond the bandwidths of liquid-crystal-on-silicon pulse shaper technology~\cite{Roelens2008}, but would be readily attainable with phase shifters utlizing the electro-optic effect, a natural choice for on-chip pulse shapers based on microring modulators. Thus, moving on chip should not only improve efficiency, but also enable the update speeds desired for memristor feedback.

Memory effects are not an exclusive feature of quantum dynamics, being also present in classical physics. Our understanding regarding quantum non-Markovian behavior has markedly increased in the last few years \cite{Capela2019,Pollock2018,Milz2019}. Therefore, moving forward with this design, the question regarding the quantumness of the time-correlations generated in the output beam should be experimentally addressed by a measure of quantum non-Markovianity or by means of a Leggett-Garg inequality~\cite{Leggett1985}. Interestingly, these two concepts are deeply linked~\cite{Souza2013}.

As demonstrated by our results, a memristor can be practically implemented in a photonic system with a frequency-entangled optical fields, thus providing a novel platform for the development of quantum circuits for simulating complex quantum systems, where the characteristic non-linear behavior of the memristor can play a major role.

\acknowledgements
The authors are grateful to Pavel Lougovski and Enrique Solano for helpful discussions.\\

The authors acknowledge support from Spanish Government PGC2018-095113-B-I00 (MCIU/AEI/FEDER, UE) and Basque Government IT986-16. The authors also acknowledge support from the projects QMiCS (820505) and OpenSuperQ (820363) of the EU Flagship on Quantum Technologies, as well as the EU FET Open Grant Quromorphic. This work is supported by the U.S. Department of Energy, Office of Science, Office of Advanced Scientific Computing Research (ASCR) quantum algorithm teams program, under field work proposal number ERKJ333. LCC would like to acknowledge the financial support from the Brazilian ministries MEC and MCTIC, funding agency CNPq, and the Brazilian National Institute of Science and Technology of Quantum Information (INCT-IQ). This study was financed in part by the Coordena\c{c}\~{a}o de Aperfei\c{c}oamento de Pessoal de N\'{i}vel Superior -- Brasil (CAPES) -- Finance Code 001. A portion of this work was performed at Oak Ridge National Laboratory, managed by UT-Battelle, LLC, for the US Department of Energy under contract no. DE-AC05-00OR22725.

\end{document}